\documentstyle[12pt]{article}
\input epsf.sty
 1

\catcode`\@=11 
\makeatletter
\def\@seccntformat#1{\csname the#1\endcsname.\hskip 1em}

\makeatother
\begin{document}

\thispagestyle{empty}
\begin{flushright}

{\footnotesize\renewcommand{\baselinestretch}{.75}
           SLAC-PUB-7940\\
           September 1998\\
}
\end{flushright}

\vspace {0.5cm}

\begin{center}
{\large \bf  QCD TESTS USING $b\bar{b}g$ EVENTS\\
AND A NEW MEASUREMENT OF THE B-HADRON\\
ENERGY DISTRIBUTION$^*$}

\vspace {1.0cm}

 {\bf David Muller}

\vspace {0.2cm}

 {\bf Representing The SLD Collaboration$^{**}$}

\vspace {0.2cm}

Stanford Linear Accelerator Center \\
Stanford University, Stanford, CA~94309 \\

\vspace{1.3cm}
{\bf Abstract}
\end{center}

\renewcommand{\baselinestretch}{1.2}

We present new studies of 3-jet final states from hadronic $Z^0$ decays
recorded by the SLD experiment, in which jets are identified as quark,
antiquark or gluon.
Our gluon energy spectrum, measured over the full kinematic range,
is consistent with the predictions of QCD, and we derive a limit on an anomalous
chromomagnetic $bbg$ coupling.
We measure the parity violation in $Z^0$ decays into $b\bar{b}g$ to be
consistent with the predictions of electroweak theory and QCD,
and perform new tests of T- and CP-conservation at the $bbg$ vertex.
We also present a new technique for reconstructing the energy of a $B$ hadron
using the set of charged tracks attached to a secondary vertex.  The $B$ hadron
energy spectrum is measured over the full kinematic range,
allowing improved tests of predictions for the shape of the spectrum.
The average scaled energy is measured to be
$<\! x_B \!> = 0.719 \pm 0.005 (stat.)$ (Preliminary).

\vspace{1.3cm}
\begin{center}
{\it Presented at the XXIX$^{th}$ International Conference on High
Energy Physics,\\
23--29 July 1998, Vancouver, Canada.}
\end{center}

\vfil

\noindent
$^*$Work supported in part by Department of Energy contract DE-AC03-76SF00515.
\eject

\section{Introduction}

Experimental studies of the structure of events containing three
hadronic jets in $e^+e^-$ annihilation
have been limited by difficulties in identifying which jet is
due to the quark, which to the antiquark and which to the gluon.
Since the gluon is expected to be the lowest-energy jet in most events, the
predictions of QCD have been tested using energy and angle distributions of
energy-ordered jets, and this is sufficient to confirm the $q\bar{q}g$ origin of
such events and to determine the gluon spin \cite{gspin}.
Tagging of the origin of any two of the three jets in such events would allow
more complete and stringent tests of QCD predictions.

Here we present a study \cite{conf1}
of 3-jet final states in which two of the jets have been
tagged as $b$ or $\bar{b}$ jets using the long lifetime of the
$B$-hadrons in the jets and the precision vertexing system of the SLD.
The remaining jet is tagged as the gluon jet, and its energy spectrum is studied
over its full kinematic range.
Adding a tag of the charge of one of the $b$/$\bar{b}$ jets,
and exploiting the high electron beam polarization of the SLC,
we measure two angular asymmetries sensitive to parity violation in the
$Z^0$ decay, and also construct new tests of T- and CP-conservation at the
$bbg$ vertex.
The study of $b$-flavor events is especially useful as input to measurements
of electroweak parameters such as $R_b$ and $A_b$,
and as a probe of new physics, which is expected in many cases to couple
more strongly to heavier quarks.
A test of the flavor-independence of the strong coupling using 3-jet final
states was presented separately \cite{conf2} at this conference.

Experimental studies of the fragmentation of bottom quarks into $B$ hadrons have
been limited by the efficiency for reconstructing the energies of individual $B$
hadrons with good resolution, especially for low energy $B$ hadrons.
Improved measurements would allow stringent tests of QCD predictions for the
fragmentation
function as well as provide crucial input to fragmentation models used for a
number of important electroweak and heavy flavor measurements.

Here we present a study \cite{conf3} of the $B$-hadron energy distribution
using a novel kinematic technique to estimate the energy $E_B$ of each
individual $B$ hadron.
$B$ hadrons are again tagged using their long lifetime, by
reconstructing secondary vertices.  The tracks in each vertex are used, along
with the vertex flight direction, to compute kinematic quantities that are then
used to select a final sample and estimate each $E_B$.
This technique has high efficiency and good energy resolution for all $E_B$, and
the resulting measurement covers the full kinematic range.

\section{The Gluon Energy Spectrum}

Well contained hadronic events \cite{evsel} in which exactly 3 jets
are found using the JADE algorithm at $y_{cut}=0.02$ are selected.
The jet energies are calculated from the angles between them \cite{evsel} and
the jets are energy ordered such that $E_1>E_2>E_3$.
In each jet we count the number $n_{sig}$ of `significant' tracks, i.e.
those with
normalized transverse impact parameter with respect to the primary interaction
point (IP) $d/\sigma_d > 3$.
We require exactly two of the three jets to have $n_{sig} > 1$, and the
remaining jet is tagged as the gluon jet.
This yields 1533 events with an estimated purity of correctly tagged gluon
jets of 91\%.
In 2.5\% (12.5\%) of these events, jet 1(2), the (second) highest energy jet,
is tagged as the gluon jet, giving coverage over the full kinematic range.

\begin{figure}
   \epsfxsize=5.2in
   \begin{center}\mbox{\epsffile{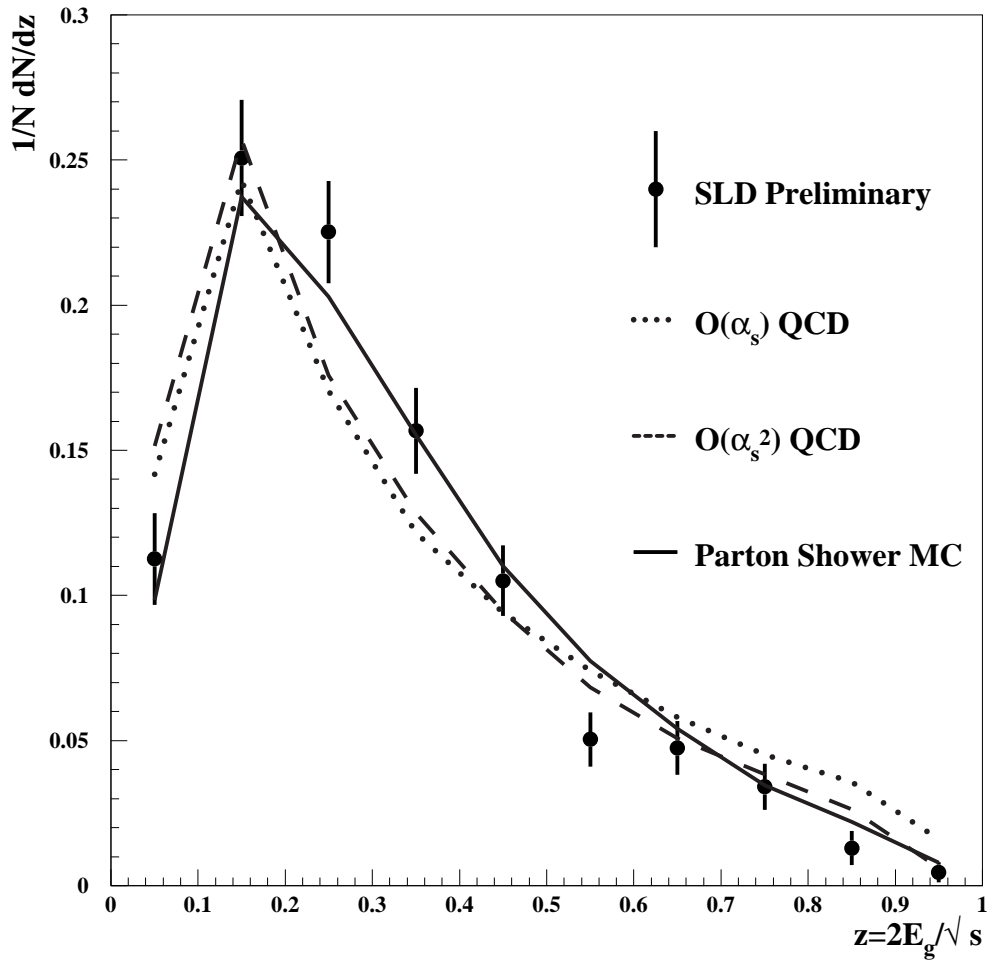}}\end{center}
\caption{ \label{xg}
The measured scaled gluon energy distribution (dots).
The predictions of first and second order QCD
and of a parton shower calculation are also shown as the dotted, dashed and
solid lines, respectively.
  }
\end{figure}

The background from non-$b\bar{b}g$ events and events with an incorrect gluon
tag is subtracted, and the resulting distribution of scaled gluon energy
$z=2E_g/\sqrt{s}$ is corrected for the effects of selection efficiency and
resolution.
The fully corrected spectrum is shown in fig. \ref{xg}, and shows the
expected falling behaviour with increasing $z$.
The distribution is cut off at low $z$ by the finite $y_{cut}$ value used
for jet finding.
Also shown are the predictions of first and second order QCD.
Both reproduce the general behaviour, but fail to describe the data in detail.
The prediction of the JETSET \cite{jetset} parton shower simulation is also
shown and reproduces the data.
Our data thus confirm the predictions of QCD,
although higher order effects are clearly important in the intermediate gluon
energy range, $0.2<z<0.4$.

The gluon energy spectrum is particularly sensitive
to the presence of an anomalous chromomagnetic term in the QCD Lagrangian.
A fit of the theoretical prediction \cite{rizzo} including an anomalous
term parametrized by a relative coupling $\kappa$, yields a value of
$\kappa = -0.03 \pm 0.06$ (Preliminary), consistent with zero, and
corresponding to limits on such contributions to the $bbg$ coupling of
$-0.15<\kappa <0.09$ at the 95\% confidence level.

\begin{figure}
   \epsfxsize=7.1in
   \begin{center}\mbox{\epsffile{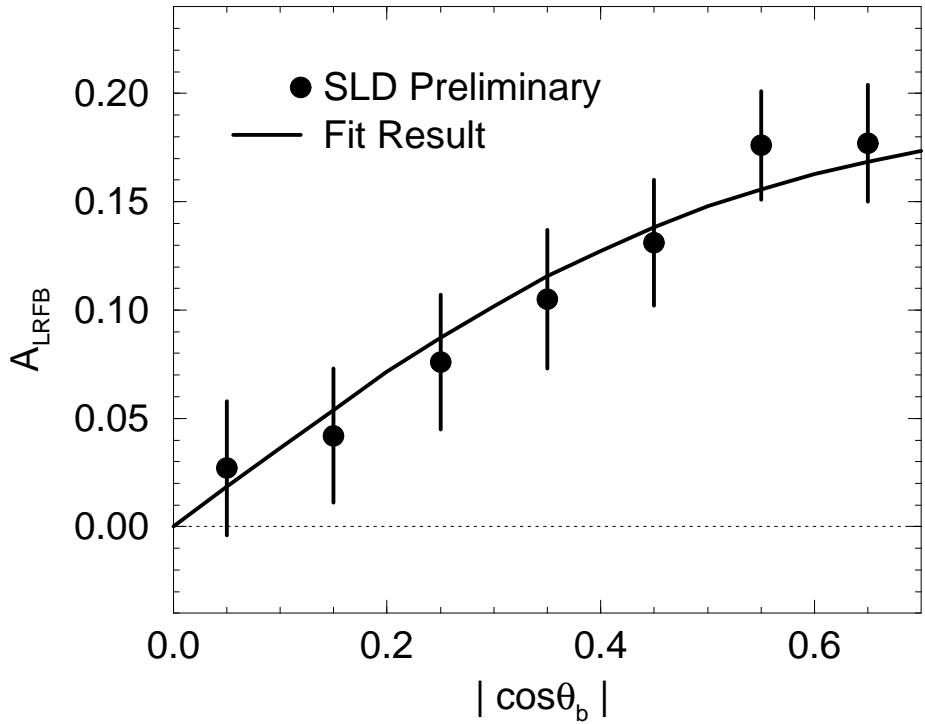}}\end{center}
\caption{ \label{pviol}
Left-right-forward-backward asymmetry of the $b$-quark
polar angle in 3-jet $Z^0$ decays.
The line is the result of a fit.}
\end{figure}

\section{Parity Violation in 3-jet $Z^0$ Decays}

We now consider two angles, the polar angle of the quark with respect to
the electron beam direction $\theta_q$, and
the angle between the quark-gluon and quark-electron beam planes
$\chi=\cos^{-1} (\hat{p}_q \times \hat{p}_g) \cdot (\hat{p}_q \times
\hat{p}_e)$.
The cosine $x$ of each of these angles is expected to be distributed as
$1+x^2+2 A_P A_Z x$, where the $Z^0$ polarization $A_Z = (P_e-A_e)/(1-P_eA_e)$
depends on the electron beam polarization $P_e$, and each
$A_P$ is predicted by QCD.

Three-jet events (Durham algorithm, $y_{cut}=0.005$) are selected and energy
ordered,
and a topological vertex finder \cite{dj} is applied to the tracks in each jet.
The 3420 events containing any vertex with invariant
mass above 1.5 GeV/c$^2$ are
kept; these have an estimated $b\bar{b}g$ purity of 87\%.
We calculate the momentum-weighted charge of each jet $j$,
$Q_j=\Sigma_i q_i |\vec{p}_i \cdot \hat{p}_j |^{0.5}$, using the charge $q_i$
and momentum $\vec{p}_i$ of each track $i$ in the jet.
In this case we assume that the highest-energy jet is not the gluon, and
tag it as the $b$ ($\bar{b}$) jet if $Q = Q_1 - Q_2 - Q_3$ is negative
(positive).
We then define the $b$-quark polar angle
$\cos\theta_b = -{\rm sign}(Q)(\hat{p}_e \cdot \hat{p}_1)$.

The left-right-forward-backward asymmetry $\tilde{A}_{FB}^b$ in
this polar angle is shown as a function of $|\cos\theta_b|$ in
fig. \ref{pviol}.
A clear asymmetry is seen, which increases with $|\cos\theta_b|$ in the
expected way.
A fit yields an asymmetry parameter of $A_P = 0.89 \pm 0.06 \pm 0.07$
(Preliminary), where the first error is statistical and the second
systematic, consistent with the QCD prediction of
$A_P=0.93A_b = 0.87$.

\begin{figure}
   \epsfxsize=6.8in
   \begin{center}\mbox{\epsffile{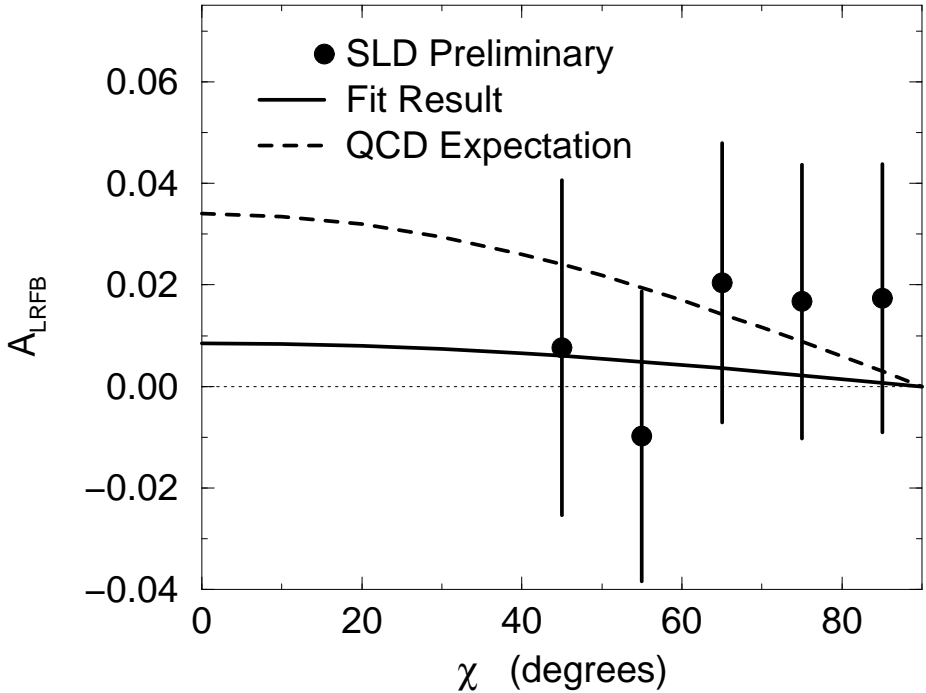}}\end{center}
\caption{ \label{pvchi}
Left-right-forward-backward asymmetry of the angle $\chi$ in 3-jet $Z^0$ decays.
The dashed and solid lines are the QCD prediction and the result of a fit,
respectively.}
\end{figure}

We then tag one of the two lower energy jets as the gluon jet, using the
impact parameters of their tracks.
If jet 2 has $n_{sig}=0$ and jet 3 has $n_{sig}>0$, then jet 2 is tagged as
the gluon jet; otherwise jet 3 is tagged as the gluon jet.
In each event we construct the angle $\chi$, and $\tilde{A}_{FB}^{\chi}$ is
shown as a function of $\chi$ in fig. \ref{pvchi}.
Here we expect only a small deviation from zero as indicated by the dashed
line on fig. \ref{pvchi}.
Our measurement is consistent with the prediction, as well as with zero.
A fit (solid line) yields
$A_{\chi} = -0.015 \pm 0.045 \pm 0.001$
(Preliminary), to be compared with an expectation of $-$0.060.

\section{Symmetry Tests in 3-jet $Z^0$ Decays}

Using these fully tagged events, we can construct observables that are formally
odd under time and/or CP reversal.
The energy-ordered triple product 
$\cos\omega^+ = \vec{\sigma}_Z \cdot (\hat{p}_1 \times \hat{p}_2)$, where
$\vec{\sigma}_Z$ is the $Z^0$ polarization vector, is $T_N$-odd and CP-even.
Since the true time reversed experiment is not performed, this
quantity could have a nonzero $\tilde{A}_{FB}$,
and we have previously set a limit~\cite{evsel} using events of all flavors.
A calculation \cite{lance} including Standard Model final state interactions
predicts that $\tilde{A}_{FB}^{\omega^+}$ is largest for $b\bar{b}g$
events, but is only $\sim$10$^{-5}$.
The fully flavor-ordered triple product 
$\cos\omega^- = \vec{\sigma}_Z \cdot (\hat{p}_q \times \hat{p}_{\bar{q}})$
is both $T_N$-odd and CP-odd.

\begin{figure}
   \epsfxsize=5.8in
   \begin{center}\mbox{\epsffile{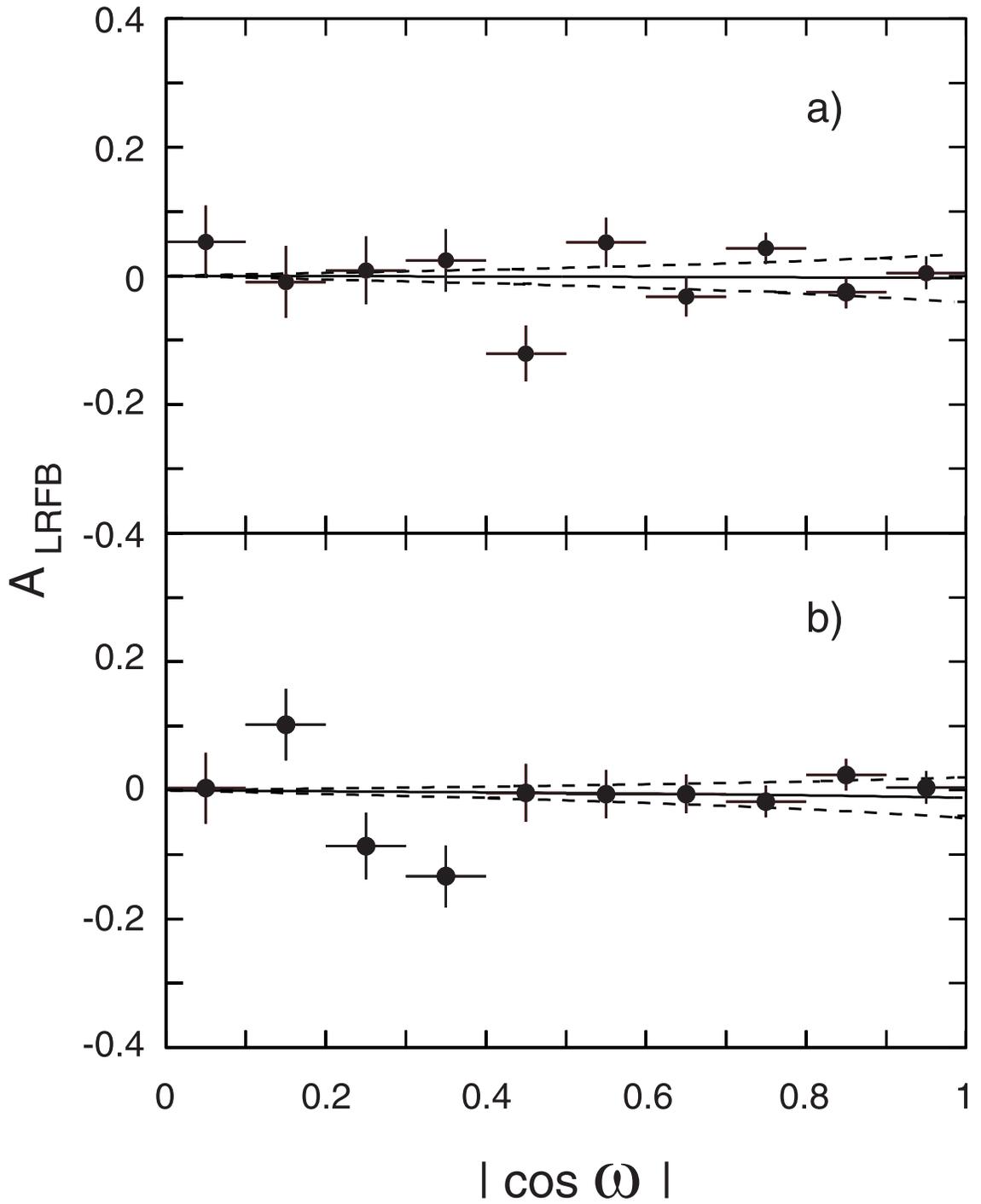}}\end{center}
\caption{ \label{tcp}
Left-right-forward-backward asymmetries of the a) energy- and
b) flavor-ordered triple product.
The solid (dashed) line represents a fit to the data (95\% confidence limits).
 }
\end{figure}

Our measured $\tilde{A}_{FB}^{\omega^+}$ and $\tilde{A}_{FB}^{\omega^-}$ are
shown in fig. \ref{tcp}.
They are consistent with zero at all $|\cos\omega|$.
Fits (solid lines in fig. \ref{tcp}) to the data yield the asymmetry parameters
consistent with zero, and we set limits on any $T_N$- and CP-violating
asymmetries of $-0.039<A^+_T<0.035$ and $-0.086<A^-_T<0.040$, respectively.

\section{The $B$-Hadron Energy Spectrum}

For the $B$ energy measurement, each selected hadronic event is divided into
two hemispheres by the plane perpendicular to the thrust axis.  The topological
vertex finder is applied to the set of tracks in each hemisphere, and any vertex
with invariant
mass greater than 2 GeV/c$^2$ is retained as a candidate $B$ hadron vertex.
The flight direction of the $B$ hadron is taken to be along the line joining the
IP and the vertex position.

The four-vector sum of the tracks in the vertex (assigned the charged pion mass)
is calculated, and its momentum component transverse to the flight direction
is equated with the transverse component of the ``missing" momentum from the $B$
hadron decay.  At this point two quantities are still needed in order to
determine the energy of the $B$ hadron, the missing mass and the missing
momentum along the flight direction.
Assuming a mass $M_B$ for the $B$ hadron eliminates one of these unknowns,
and also allows an upper limit to be calculated on the missing mass:
\begin{equation}
M^2_{0max} = M^2_B - 2M_B \sqrt{M^2_{chg} + P^2_t} + M^2_{chg},
\end{equation}
where $M_{chg}$ and $P_t$ are the invariant mass and momentum transverse to the
flight direction, respectively, of the set of tracks in the vertex.
Using $M_B=5.28$ GeV/c$^2$, equating $M^2_0$ with $M^2_{0max}$ and solving for
the $B$ hadron energy is found to provide a good estimate of the true $B$
hadron energy for the mixture of $B$ hadron species produced in our simulation
of $Z^0$ decays.
As expected, the simulated resolution on this estimate is best for vertices
with low values of $M^2_{0max}$, approaching 6\% as $M^2_{0max} \rightarrow 0$.
It does not degrade very rapidly with increasing $M^2_{0max}$
due to the strong tendency for the true missing mass in hadronic $B$ decays to
cluster near the maximum value.

\begin{figure}
   \epsfxsize=7.3in
   \begin{center}\mbox{\epsffile{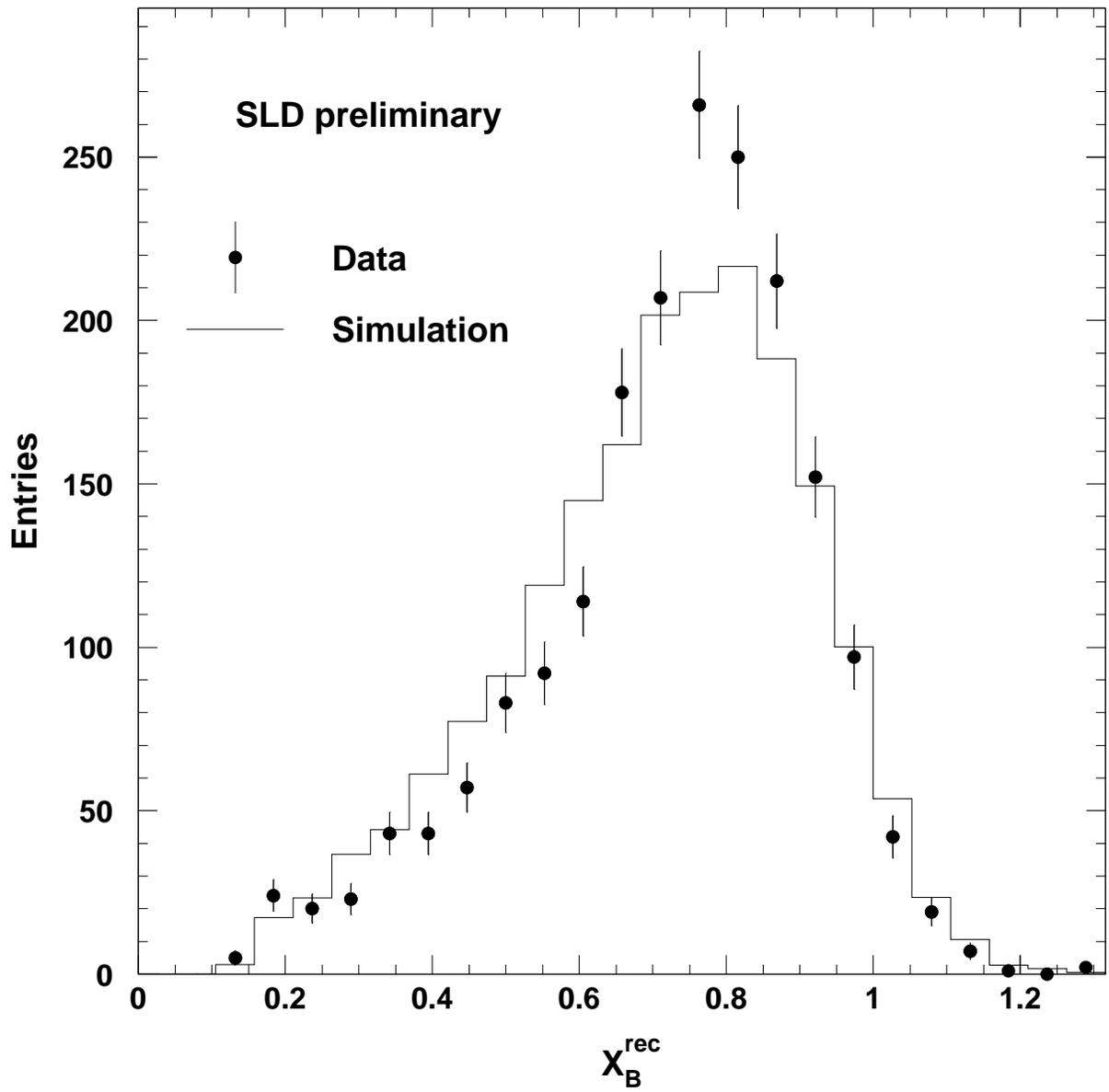}}\end{center}
\caption{ \label{ebfig}
Uncorrected distribution of measured $B$ hadron energies (dots).
The histogram is the prediction of our simulation.}
\end{figure}

A cut is placed on $M^2_{0max}$ that depends on the measured energy in such a
way that the efficiency for selecting $B$ vertices, estimated from our
simulation, is roughly independent of energy;
it is 4\% on average and is above 3\% for $E_B>8$ GeV.
A sample of 1938 vertices is selected, with an estimated $B$ hadron purity of
99.5\%.
The energy resolution is estimated to be 10\% on average, roughly independent
of $E_B$.
The measured energy is then divided by the beam energy, and the distribution of
$x_B=E_B/E_{beam}$ is shown in fig. \ref{ebfig}.
It can be seen that the measurement covers the entire kinematic range from the
$B$ hadron mass ($x_B \approx 0.12$) to the beam energy.

Also shown in fig. \ref{ebfig} is the prediction of our simulation, generated
using the JETSET program with the Peterson fragmentation option and
$\epsilon_b=0.006$.  The predicted distribution peaks at a value of $E_B$
consistent with the peak in the data, but the width is significantly larger than
that of the data.

The correction of the measured distribution to obtain the true $x_B$ 
distribution depends on the form assumed for the true distribution,
due to the rapid variation of the distribution on the scale of
the bin size.
We have tested several functional forms for the true $x_B$ distribution by
weighting simulated $B$ hadrons at the generator level to reproduce the
function for a given set of parameter values.
The weighted detector-level distribution is compared with that measured in the
data, and a $\chi^2$ is minimized to find the best parameter values.
Most of the test functions we tried \cite{conf3} are not able to describe the
data adequately for any values of their parameters.
Three functions, the Peterson function and two generalizations thereof,
are able to describe the data; at their respective fitted parameter values
they are very similar to each other for
$x_B<0.7$, but show substantial variation for $x_B>0.8$, indicating that there
remains nonnegligible model-dependence in the correction.
From these three fitted functions we extract a measurement of the mean value,
\begin{equation}
<\!\! x_B \!\!> = 0.719 \pm 0.005 \;(stat.) \pm 0.001 \;(shape) \;\;\;
{\rm (Preliminary).}
\end{equation}
The experimental systematic error and the correction procedure are under
study.

\section{Conclusions}

In summary, we use the excellent vertexing capability of the SLD and
the high electron beam polarization of the SLC to make several new
tests of QCD, using 3-jet final states in which jets are tagged as quark,
antiquark or gluon jets.
The gluon energy spectrum is measured over its full kinematic range; we confirm
the prediction of QCD and set limits on anomalous chromomagnetic couplings.
The parity violation in $Z^0$ decays to $b\bar{b}g$ is found to be consistent
with the predictions of electroweak theory plus QCD, and new tests of T- and
CP-conservation in strong interactions are performed.

In addition, we have developed a new technique for measuring the energies of
individual $B$ hadrons using information only from the charged tracks attached
to a secondary vertex.
The method has high efficiency and good energy resolution that are roughly
independent of the true energy.
The $B$-hadron energy distribution has been measured over the full kinematic
range, providing a basis for improved tests of QCD predictions and better input
into fragmentation models.

\section*{Acknowledgements}
We thank the personnel of the SLAC accelerator department and the
technical
staffs of our collaborating institutions for their outstanding efforts
on our behalf.
We thank L.~Dixon and T. Rizzo for useful discussions.

%

\section*{$^{**}$List of Authors} 
%
%
%
\begin{center}
\def\iADEL{$^{(1)}$}
\def\iAOMORI{$^{(2)}$}
\def\iBOLO{$^{(3)}$}
\def\iBRUN{$^{(4)}$}
\def\iBU{$^{(5)}$}
\def\iCINC{$^{(6)}$}
\def\iCOLO{$^{(7)}$}
\def\iCOLU{$^{(8)}$}
\def\iCSU{$^{(9)}$}
\def\iFERR{$^{(10)}$}
\def\iFRAS{$^{(11)}$}
\def\iILLI{$^{(12)}$}
\def\iLBL{$^{(13)}$}
\def\iLTU{$^{(14)}$}
\def\iMASS{$^{(15)}$}
\def\iMISSI{$^{(16)}$}
\def\iMIT{$^{(17)}$}
\def\iMOSCOW{$^{(18)}$}
\def\iNAGO{$^{(19)}$}
\def\iOREG{$^{(20)}$}
\def\iOXF{$^{(21)}$}
\def\iPADO{$^{(22)}$}
\def\iPERU{$^{(23)}$}
\def\iPISA{$^{(24)}$}
\def\iRAL{$^{(25)}$}
\def\iRUTG{$^{(26)}$}
\def\iSLAC{$^{(27)}$}
\def\iSOGA{$^{(28)}$}
\def\iSOONG{$^{(29)}$}
\def\iTENN{$^{(30)}$}
\def\iTOHO{$^{(31)}$}
\def\iUCSB{$^{(32)}$}
\def\iUCSC{$^{(33)}$}
\def\iVAND{$^{(34)}$}
\def\iWASH{$^{(35)}$}
\def\iWISC{$^{(36)}$}
\def\iYALE{$^{(37)}$}

  \baselineskip=.75\baselineskip  
\mbox{K. Abe\unskip,\iAOMORI}
\mbox{K.  Abe\unskip,\iNAGO}
\mbox{T. Abe\unskip,\iSLAC}
\mbox{I.Adam\unskip,\iSLAC}
\mbox{T.  Akagi\unskip,\iSLAC}
\mbox{N. J. Allen\unskip,\iBRUN}
\mbox{A. Arodzero\unskip,\iOREG}
\mbox{W.W. Ash\unskip,\iSLAC}
\mbox{D. Aston\unskip,\iSLAC}
\mbox{K.G. Baird\unskip,\iMASS}
\mbox{C. Baltay\unskip,\iYALE}
\mbox{H.R. Band\unskip,\iWISC}
\mbox{M.B. Barakat\unskip,\iLTU}
\mbox{O. Bardon\unskip,\iMIT}
\mbox{T.L. Barklow\unskip,\iSLAC}
\mbox{J.M. Bauer\unskip,\iMISSI}
\mbox{G. Bellodi\unskip,\iOXF}
\mbox{R. Ben-David\unskip,\iYALE}
\mbox{A.C. Benvenuti\unskip,\iBOLO}
\mbox{G.M. Bilei\unskip,\iPERU}
\mbox{D. Bisello\unskip,\iPADO}
\mbox{G. Blaylock\unskip,\iMASS}
\mbox{J.R. Bogart\unskip,\iSLAC}
\mbox{B. Bolen\unskip,\iMISSI}
\mbox{G.R. Bower\unskip,\iSLAC}
\mbox{J. E. Brau\unskip,\iOREG}
\mbox{M. Breidenbach\unskip,\iSLAC}
\mbox{W.M. Bugg\unskip,\iTENN}
\mbox{D. Burke\unskip,\iSLAC}
\mbox{T.H. Burnett\unskip,\iWASH}
\mbox{P.N. Burrows\unskip,\iOXF}
\mbox{A. Calcaterra\unskip,\iFRAS}
\mbox{D.O. Caldwell\unskip,\iUCSB}
\mbox{D. Calloway\unskip,\iSLAC}
\mbox{B. Camanzi\unskip,\iFERR}
\mbox{M. Carpinelli\unskip,\iPISA}
\mbox{R. Cassell\unskip,\iSLAC}
\mbox{R. Castaldi\unskip,\iPISA}
\mbox{A. Castro\unskip,\iPADO}
\mbox{M. Cavalli-Sforza\unskip,\iUCSC}
\mbox{A. Chou\unskip,\iSLAC}
\mbox{E. Church\unskip,\iWASH}
\mbox{H.O. Cohn\unskip,\iTENN}
\mbox{J.A. Coller\unskip,\iBU}
\mbox{M.R. Convery\unskip,\iSLAC}
\mbox{V. Cook\unskip,\iWASH}
\mbox{R. Cotton\unskip,\iBRUN}
\mbox{R.F. Cowan\unskip,\iMIT}
\mbox{D.G. Coyne\unskip,\iUCSC}
\mbox{G. Crawford\unskip,\iSLAC}
\mbox{C.J.S. Damerell\unskip,\iRAL}
\mbox{M. N. Danielson\unskip,\iCOLO}
\mbox{M. Daoudi\unskip,\iSLAC}
\mbox{N. de Groot\unskip,\iSLAC}
\mbox{R. Dell'Orso\unskip,\iPERU}
\mbox{P.J. Dervan\unskip,\iBRUN}
\mbox{R. de Sangro\unskip,\iFRAS}
\mbox{M. Dima\unskip,\iCSU}
\mbox{A. D'Oliveira\unskip,\iCINC}
\mbox{D.N. Dong\unskip,\iMIT}
\mbox{P.Y.C. Du\unskip,\iTENN}
\mbox{R. Dubois\unskip,\iSLAC}
\mbox{B.I. Eisenstein\unskip,\iILLI}
\mbox{V. Eschenburg\unskip,\iMISSI}
\mbox{E. Etzion\unskip,\iWISC}
\mbox{S. Fahey\unskip,\iCOLO}
\mbox{D. Falciai\unskip,\iFRAS}
\mbox{C. Fan\unskip,\iCOLO}
\mbox{J.P. Fernandez\unskip,\iUCSC}
\mbox{M.J. Fero\unskip,\iMIT}
\mbox{K.Flood\unskip,\iMASS}
\mbox{R. Frey\unskip,\iOREG}
\mbox{T. Gillman\unskip,\iRAL}
\mbox{G. Gladding\unskip,\iILLI}
\mbox{S. Gonzalez\unskip,\iMIT}
\mbox{E.L. Hart\unskip,\iTENN}
\mbox{J.L. Harton\unskip,\iCSU}
\mbox{A. Hasan\unskip,\iBRUN}
\mbox{K. Hasuko\unskip,\iTOHO}
\mbox{S. J. Hedges\unskip,\iBU}
\mbox{S.S. Hertzbach\unskip,\iMASS}
\mbox{M.D. Hildreth\unskip,\iSLAC}
\mbox{J. Huber\unskip,\iOREG}
\mbox{M.E. Huffer\unskip,\iSLAC}
\mbox{E.W. Hughes\unskip,\iSLAC}
\mbox{X.Huynh\unskip,\iSLAC}
\mbox{H. Hwang\unskip,\iOREG}
\mbox{M. Iwasaki\unskip,\iOREG}
\mbox{D. J. Jackson\unskip,\iRAL}
\mbox{P. Jacques\unskip,\iRUTG}
\mbox{J.A. Jaros\unskip,\iSLAC}
\mbox{Z.Y. Jiang\unskip,\iSLAC}
\mbox{A.S. Johnson\unskip,\iSLAC}
\mbox{J.R. Johnson\unskip,\iWISC}
\mbox{R.A. Johnson\unskip,\iCINC}
\mbox{T. Junk\unskip,\iSLAC}
\mbox{R. Kajikawa\unskip,\iNAGO}
\mbox{M. Kalelkar\unskip,\iRUTG}
\mbox{Y. Kamyshkov\unskip,\iTENN}
\mbox{H.J. Kang\unskip,\iRUTG}
\mbox{I. Karliner\unskip,\iILLI}
\mbox{H. Kawahara\unskip,\iSLAC}
\mbox{Y. D. Kim\unskip,\iSOGA}
\mbox{R. King\unskip,\iSLAC}
\mbox{M.E. King\unskip,\iSLAC}
\mbox{R.R. Kofler\unskip,\iMASS}
\mbox{N.M. Krishna\unskip,\iCOLO}
\mbox{R.S. Kroeger\unskip,\iMISSI}
\mbox{M. Langston\unskip,\iOREG}
\mbox{A. Lath\unskip,\iMIT}
\mbox{D.W.G. Leith\unskip,\iSLAC}
\mbox{V. Lia\unskip,\iMIT}
\mbox{C.-J. S. Lin\unskip,\iSLAC}
\mbox{X. Liu\unskip,\iUCSC}
\mbox{M.X. Liu\unskip,\iYALE}
\mbox{M. Loreti\unskip,\iPADO}
\mbox{A. Lu\unskip,\iUCSB}
\mbox{H.L. Lynch\unskip,\iSLAC}
\mbox{J. Ma\unskip,\iWASH}
\mbox{G. Mancinelli\unskip,\iRUTG}
\mbox{S. Manly\unskip,\iYALE}
\mbox{G. Mantovani\unskip,\iPERU}
\mbox{T.W. Markiewicz\unskip,\iSLAC}
\mbox{T. Maruyama\unskip,\iSLAC}
\mbox{H. Masuda\unskip,\iSLAC}
\mbox{E. Mazzucato\unskip,\iFERR}
\mbox{A.K. McKemey\unskip,\iBRUN}
\mbox{B.T. Meadows\unskip,\iCINC}
\mbox{G. Menegatti\unskip,\iFERR}
\mbox{R. Messner\unskip,\iSLAC}
\mbox{P.M. Mockett\unskip,\iWASH}
\mbox{K.C. Moffeit\unskip,\iSLAC}
\mbox{T.B. Moore\unskip,\iYALE}
\mbox{M.Morii\unskip,\iSLAC}
\mbox{D. Muller\unskip,\iSLAC}
\mbox{V.Murzin\unskip,\iMOSCOW}
\mbox{T. Nagamine\unskip,\iTOHO}
\mbox{S. Narita\unskip,\iTOHO}
\mbox{U. Nauenberg\unskip,\iCOLO}
\mbox{H. Neal\unskip,\iSLAC}
\mbox{M. Nussbaum\unskip,\iCINC}
\mbox{N.Oishi\unskip,\iNAGO}
\mbox{D. Onoprienko\unskip,\iTENN}
\mbox{L.S. Osborne\unskip,\iMIT}
\mbox{R.S. Panvini\unskip,\iVAND}
\mbox{H. Park\unskip,\iOREG}
\mbox{C. H. Park\unskip,\iSOONG}
\mbox{T.J. Pavel\unskip,\iSLAC}
\mbox{I. Peruzzi\unskip,\iFRAS}
\mbox{M. Piccolo\unskip,\iFRAS}
\mbox{L. Piemontese\unskip,\iFERR}
\mbox{E. Pieroni\unskip,\iPISA}
\mbox{K.T. Pitts\unskip,\iOREG}
\mbox{R.J. Plano\unskip,\iRUTG}
\mbox{R. Prepost\unskip,\iWISC}
\mbox{C.Y. Prescott\unskip,\iSLAC}
\mbox{G.D. Punkar\unskip,\iSLAC}
\mbox{J. Quigley\unskip,\iMIT}
\mbox{B.N. Ratcliff\unskip,\iSLAC}
\mbox{T.W. Reeves\unskip,\iVAND}
\mbox{J. Reidy\unskip,\iMISSI}
\mbox{P.L. Reinertsen\unskip,\iUCSC}
\mbox{P.E. Rensing\unskip,\iSLAC}
\mbox{L.S. Rochester\unskip,\iSLAC}
\mbox{P.C. Rowson\unskip,\iCOLU}
\mbox{J.J. Russell\unskip,\iSLAC}
\mbox{O.H. Saxton\unskip,\iSLAC}
\mbox{T. Schalk\unskip,\iUCSC}
\mbox{R.H. Schindler\unskip,\iSLAC}
\mbox{B.A. Schumm\unskip,\iUCSC}
\mbox{J. Schwiening\unskip,\iSLAC}
\mbox{S. Sen\unskip,\iYALE}
\mbox{V.V. Serbo\unskip,\iWISC}
\mbox{M.H. Shaevitz\unskip,\iCOLU}
\mbox{J.T. Shank\unskip,\iBU}
\mbox{G. Shapiro\unskip,\iLBL}
\mbox{D.J. Sherden\unskip,\iSLAC}
\mbox{K. D. Shmakov\unskip,\iTENN}
\mbox{C. Simopoulos\unskip,\iSLAC}
\mbox{N.B. Sinev\unskip,\iOREG}
\mbox{S.R. Smith\unskip,\iSLAC}
\mbox{M. B. Smy\unskip,\iCSU}
\mbox{J.A. Snyder\unskip,\iYALE}
\mbox{H. Staengle\unskip,\iCSU}
\mbox{A. Stahl\unskip,\iSLAC}
\mbox{P. Stamer\unskip,\iRUTG}
\mbox{R. Steiner\unskip,\iADEL}
\mbox{H. Steiner\unskip,\iLBL}
\mbox{M.G. Strauss\unskip,\iMASS}
\mbox{D. Su\unskip,\iSLAC}
\mbox{F. Suekane\unskip,\iTOHO}
\mbox{A. Sugiyama\unskip,\iNAGO}
\mbox{S. Suzuki\unskip,\iNAGO}
\mbox{M. Swartz\unskip,\iSLAC}
\mbox{A. Szumilo\unskip,\iWASH}
\mbox{T. Takahashi\unskip,\iSLAC}
\mbox{F.E. Taylor\unskip,\iMIT}
\mbox{J. Thom\unskip,\iSLAC}
\mbox{E. Torrence\unskip,\iMIT}
\mbox{N. K. Toumbas\unskip,\iSLAC}
\mbox{A.I. Trandafir\unskip,\iMASS}
\mbox{J.D. Turk\unskip,\iYALE}
\mbox{T. Usher\unskip,\iSLAC}
\mbox{C. Vannini\unskip,\iPISA}
\mbox{J. Va'vra\unskip,\iSLAC}
\mbox{E. Vella\unskip,\iSLAC}
\mbox{J.P. Venuti\unskip,\iVAND}
\mbox{R. Verdier\unskip,\iMIT}
\mbox{P.G. Verdini\unskip,\iPISA}
\mbox{S.R. Wagner\unskip,\iSLAC}
\mbox{D. L. Wagner\unskip,\iCOLO}
\mbox{A.P. Waite\unskip,\iSLAC}
\mbox{Walston, S.\unskip,\iOREG}
\mbox{J.Wang\unskip,\iSLAC}
\mbox{C. Ward\unskip,\iBRUN}
\mbox{S.J. Watts\unskip,\iBRUN}
\mbox{A.W. Weidemann\unskip,\iTENN}
\mbox{E. R. Weiss\unskip,\iWASH}
\mbox{J.S. Whitaker\unskip,\iBU}
\mbox{S.L. White\unskip,\iTENN}
\mbox{F.J. Wickens\unskip,\iRAL}
\mbox{B. Williams\unskip,\iCOLO}
\mbox{D.C. Williams\unskip,\iMIT}
\mbox{S.H. Williams\unskip,\iSLAC}
\mbox{S. Willocq\unskip,\iSLAC}
\mbox{R.J. Wilson\unskip,\iCSU}
\mbox{W.J. Wisniewski\unskip,\iSLAC}
\mbox{J. L. Wittlin\unskip,\iMASS}
\mbox{M. Woods\unskip,\iSLAC}
\mbox{G.B. Word\unskip,\iVAND}
\mbox{T.R. Wright\unskip,\iWISC}
\mbox{J. Wyss\unskip,\iPADO}
\mbox{R.K. Yamamoto\unskip,\iMIT}
\mbox{J.M. Yamartino\unskip,\iMIT}
\mbox{X. Yang\unskip,\iOREG}
\mbox{J. Yashima\unskip,\iTOHO}
\mbox{S.J. Yellin\unskip,\iUCSB}
\mbox{C.C. Young\unskip,\iSLAC}
\mbox{H. Yuta\unskip,\iAOMORI}
\mbox{G. Zapalac\unskip,\iWISC}
\mbox{R.W. Zdarko\unskip,\iSLAC}
\mbox{J. Zhou\unskip.\iOREG}

\it
  \vskip \baselineskip                   
  \centerline{(The SLD Collaboration)}   
  \vskip \baselineskip        
  \baselineskip=.75\baselineskip   
\iADEL
  Adelphi University,
  South Avenue-   Garden City,NY 11530, \break
\iAOMORI
  Aomori University,
  2-3-1 Kohata, Aomori City, 030 Japan, \break
\iBOLO
  INFN Sezione di Bologna,
  Via Irnerio 46    I-40126 Bologna  (Italy), \break
\iBRUN
  Brunel University,
  Uxbridge, Middlesex - UB8 3PH United Kingdom, \break
\iBU
  Boston University,
  590 Commonwealth Ave. - Boston,MA 02215, \break
\iCINC
  University of Cincinnati,
  Cincinnati,OH 45221, \break
\iCOLO
  University of Colorado,
  Campus Box 390 - Boulder,CO 80309, \break
\iCOLU
  Columbia University,
  Nevis Laboratories  P.O.Box 137 - Irvington,NY 10533, \break
\iCSU
  Colorado State University,
  Ft. Collins,CO 80523, \break
\iFERR
  INFN Sezione di Ferrara,
  Via Paradiso,12 - I-44100 Ferrara (Italy), \break
\iFRAS
  Lab. Nazionali di Frascati,
  Casella Postale 13   I-00044 Frascati (Italy), \break
\iILLI
  University of Illinois,
  1110 West Green St.  Urbana,IL 61801, \break
\iLBL
  Lawrence Berkeley Laboratory,
  Dept.of Physics 50B-5211 University of California-  Berkeley,CA 94720, \break
\iLTU
  Louisiana Technical University,
  , \break
\iMASS
  University of Massachusetts,
  Amherst,MA 01003, \break
\iMISSI
  University of Mississippi,
  University,MS 38677, \break
\iMIT
  Massachusetts Institute of Technology,
  77 Massachussetts Avenue  Cambridge,MA 02139, \break
\iMOSCOW
  Moscow State University,
  Institute of Nuclear Physics  119899 Moscow  Russia, \break
\iNAGO
  Nagoya University,
  Nagoya 464 Japan, \break
\iOREG
  University of Oregon,
  Department of Physics  Eugene,OR 97403, \break
\iOXF
  Oxford University,
  Oxford, OX1 3RH, United Kingdom, \break
\iPADO
  Universita di Padova,
  Via F. Marzolo,8   I-35100 Padova (Italy), \break
\iPERU
  Universita di Perugia, Sezione INFN,
  Via A. Pascoli  I-06100 Perugia (Italy), \break
\iPISA
  INFN, Sezione di Pisa,
  Via Livornese,582/AS  Piero a Grado  I-56010 Pisa (Italy), \break
\iRAL
  Rutherford Appleton Laboratory,
  Chiton,Didcot - Oxon OX11 0QX United Kingdom, \break
\iRUTG
  Rutgers University,
  Serin Physics Labs  Piscataway,NJ 08855-0849, \break
\iSLAC
  Stanford Linear Accelerator Center,
  2575 Sand Hill Road  Menlo Park,CA 94025, \break
\iSOGA
  Sogang University,
  Ricci Hall  Seoul, Korea, \break
\iSOONG
  Soongsil University,
  Dongjakgu Sangdo 5 dong 1-1    Seoul, Korea 156-743, \break
\iTENN
  University of Tennessee,
  401 A.H. Nielsen Physics Blg.  -  Knoxville,Tennessee 37996-1200, \break
\iTOHO
  Tohoku University,
  Bubble Chamber Lab. - Aramaki - Sendai 980 (Japan), \break
\iUCSB
  U.C. Santa Barbara,
  3019 Broida Hall  Santa Barbara,CA 93106, \break
\iUCSC
  U.C. Santa Cruz,
  Santa Cruz,CA 95064, \break
\iVAND
  Vanderbilt University,
  Stevenson Center,Room 5333  P.O.Box 1807,Station B  Nashville,TN 37235,
\break
\iWASH
  University of Washington,
  Seattle,WA 98105, \break
\iWISC
  University of Wisconsin,
  1150 University Avenue  Madison,WS 53706, \break
\iYALE
  Yale University,
  5th Floor Gibbs Lab. - P.O.Box 208121 - New Haven,CT 06520-8121. \break
\rm
%

\end{center}

\hfill
\end{document}